\documentclass{iaus}
\usepackage{graphicx}

\title[Planetary transits and stellar variability]{Comparing the performance of stellar variability filters for the detection of planetary transits} %% give here short title %%

\author[ A. S. Bonomo \& A. F. Lanza]   %% give here short author list %%
{ A. S. Bonomo$^1$$^,$$^2$
 \and A. F. Lanza$^2$}

\affiliation{$^1$Dipartimento di Fisica e Astronomia, Universit\`a degli Studi di Catania, Via S.~Sofia, 78,
95123, Catania, Italy \\ email: {\tt aldo.bonomo@oact.inaf.it} \\[\affilskip]
$^2$INAF - Osservatorio Astrofisico di Catania, Via S. Sofia, 78, 95123, Catania, Italy \\email: {\tt nuccio.lanza@oact.inaf.it}}

\pubyear{2008}
\volume{253}  %% insert here IAU Symposium No.
\pagerange{xxx - xxx}
\jname{Transiting Planets}
\editors{A.C. Editor, B.D. Editor \& C.E. Editor, eds.}
\begin{document}

\maketitle

\begin{abstract}
We have developed a new method to improve the transit detection of Earth-sized planets in front of solar-like stars by fitting stellar microvariability by means of a spot model. A large Monte Carlo numerical experiment has been designed to test the performance of our approach in comparison with other variability filters and fitting techniques for stars of different magnitudes and planets of different radius and orbital period, as observed by the space missions CoRoT and Kepler. Here we report on the results of this experiment.
\end{abstract}

\firstsection % if your document starts with a section,
\section{Introduction}
\noindent
We present a comparison among the performance of three methods applied to filter stellar variability for the detection of Earth-like planetary transits in the light curves of solar-like stars. This requires two steps: first, the filtering of stellar variability to remove the effects of photospheric cool spots and bright faculae, whose visibility is modulated by stellar rotation; secondly, the search for transits in the filtered light curves by means of suitable detection algorithms. 

We recently proposed a filtering method based on a model of the flux variations of the Sun as a star, the so-called 3-spot model (\cite[Lanza et al. 2003]{Lanzaetal2003}). Its performance was compared with that of another method, the 200-harmonic fitting, by \cite[Bonomo \& Lanza (2008)]{BonomoLanza2008}. They showed that the 3-spot model has a better performance than the latter when the standard deviation of the noise is at least 2-4 times larger than the central depth of the transit. On the other hand, the 200-harmonic fitting is better when the standard deviation of the noise is comparable to the transit depth.

Here we extend the comparison to the iterative non-linear filter by \cite[Aigrain \& Irwin (2004)]{AigrainIrwin2004}. A comparison among different variability filters is important since only the coupling of the best filtering method with the best planetary transit detection algorithm allows us to maximize transit detection efficiency. This is especially relevant when we want to detect small, terrestrial planets, which is a challenge to CoRoT and Kepler missions.

\eject
\section{Filtering methods}
\noindent
The filtering methods we want to compare are:
\begin{enumerate}
\item[a)] 200-harmonic fitting (\cite[Moutou et al. 2005]{Moutouetal2005}, team 3): it fits stellar variability by means of a linear combination of 200 harmonic functions whose frequencies are multiples of  the fundamental frequency $f_L=1/2T$, where $T$ is the whole duration of the time series, i. e. $T\sim150$ days in the case of the CoRoT mission; 
\item[b)] 3-spot model (\cite[Lanza et al. 2003]{Lanzaetal2003}, \cite[2007]{Lanzaetal2007}): it is a simplified physical model of solar-like variability based on the rotational modulation of the flux produced by three active regions, containing both cool spots and warm faculae, plus a constant component to account for uniformly distributed active regions. In the case of the Sun, the model accounts for the flux variability up to a time scale of 14 days, after which the position and areas of the three regions and the uniform component have to be changed; 
\item[c)] Iterative non-linear (INL) filter (\cite[Aigrain \& Irwin 2004]{AigrainIrwin2004}; \cite[Moutou et al. 2005]{Moutouetal2005}, team 5): it is based on the computation of a “continuum” by applying a sliding median-boxcar filter. Points where the difference between the continuum and the original light curve is greater than 3 standard deviations are flagged and the continuum is recomputed without the flagged points, iterating the process up to convergence. The final continuum is then subtracted from the original light curve.
\end{enumerate}

\section{Light curve simulation and analysis}
\noindent
We apply a Monte Carlo approach by simulating a large number of light curves of duration 150 days (the extension of the CoRoT long runs) for different values of planetary radius $R$ ranging from 1.0 to 2.0 Earth radii, orbital period $P$ between 5 and 50 days, and standard deviation of the photon shot noise $\sigma$ from 100 to 1000 parts per millions (ppm). A noise level $\sigma=100$ ppm is obtained for a star of $V\sim12$ observed in white light by CoRoT with 1 hr integration time, while $\sigma=200$, 300 and 1000 ppm corrispond to stars of $V\sim13$, 14 and 16, respectively, observed with the same instrument and 1 hr integration time.
The phase of the first transit is taken from a uniform random distribution. The star is assumed to have the solar radius and mass. We add stellar variability, assumed in all the cases to be given by the Total Solar Irradiance variations as observed close to the maximum of solar cycle 23 (e.g., \cite[Fr\"ohlich \& Lean 2004]{FrohlichLean2004}). For each set of planetary parameters and noise level, we simulate 100 light curves with different noise and activity realizations, for a total of 8000 light curves.

After filtering solar variability with the three different filtering methods, transits are searched by means of the BLS algorithm. The ratio of the transit depth to the noise level is indicated by $\alpha$, whose statistics determine the confidence level of a given transit detection (see 
\cite[Kov\'acs et al. 2002]{Kovacsetal2002}).

Transitless light curves are analysed in the same way to establish the transit detection threshold for each filtering method, by requiring a maximum false-alarm rate of 1 percent.

\begin{figure}
\centerline{
% \vspace*{-2.0 cm}
\begin{minipage}[t]{5.5cm}
\begin{center}
 \includegraphics[width=4.2cm,angle=90]{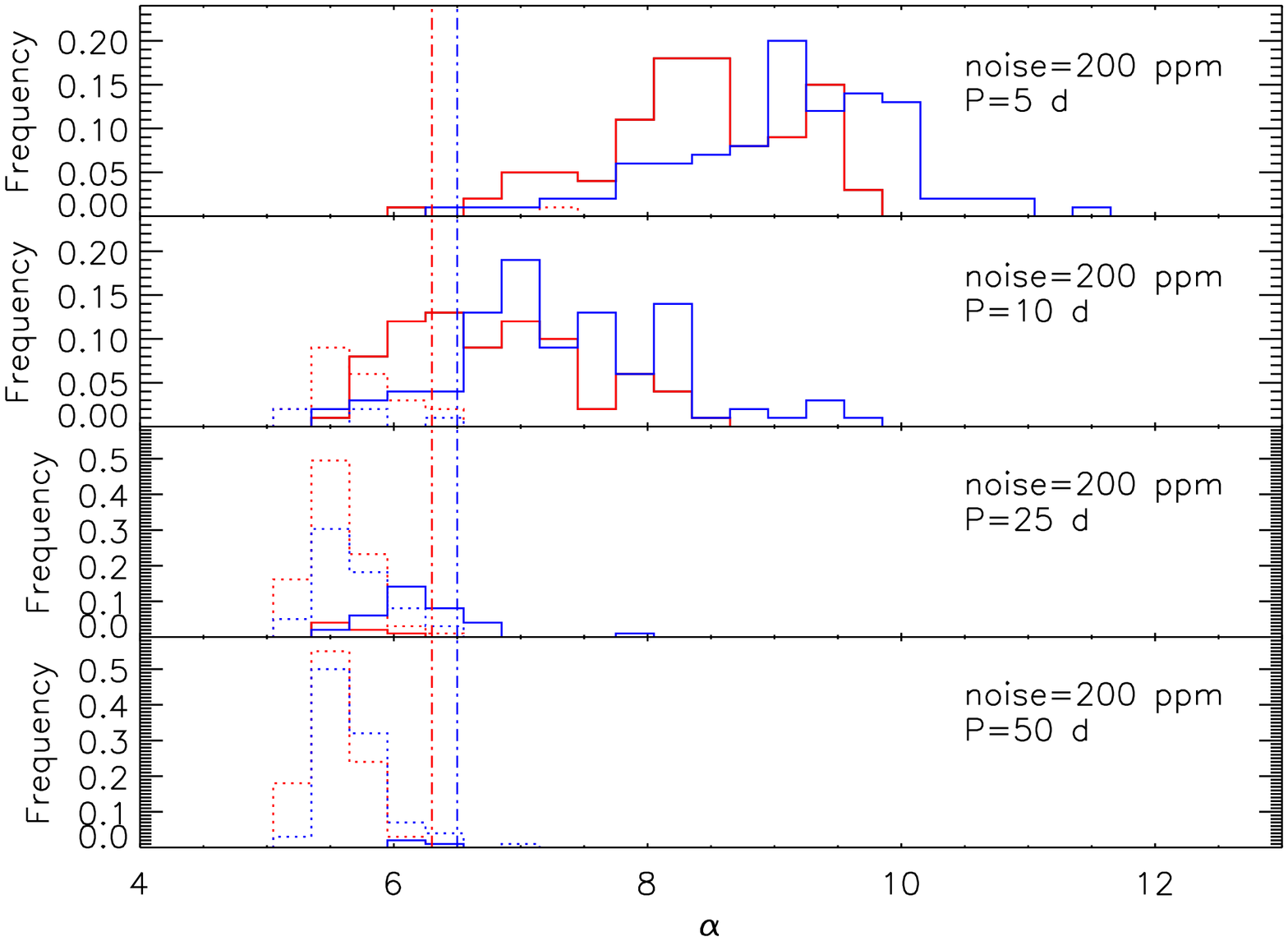} 
% \vspace*{-1.0 cm}  height=5cm,
\end{center}
\end{minipage}
%\hfill
\hspace*{1cm}
\begin{minipage}[t]{5.5cm}
\begin{center}
 \includegraphics[width=4.2cm,angle=90]{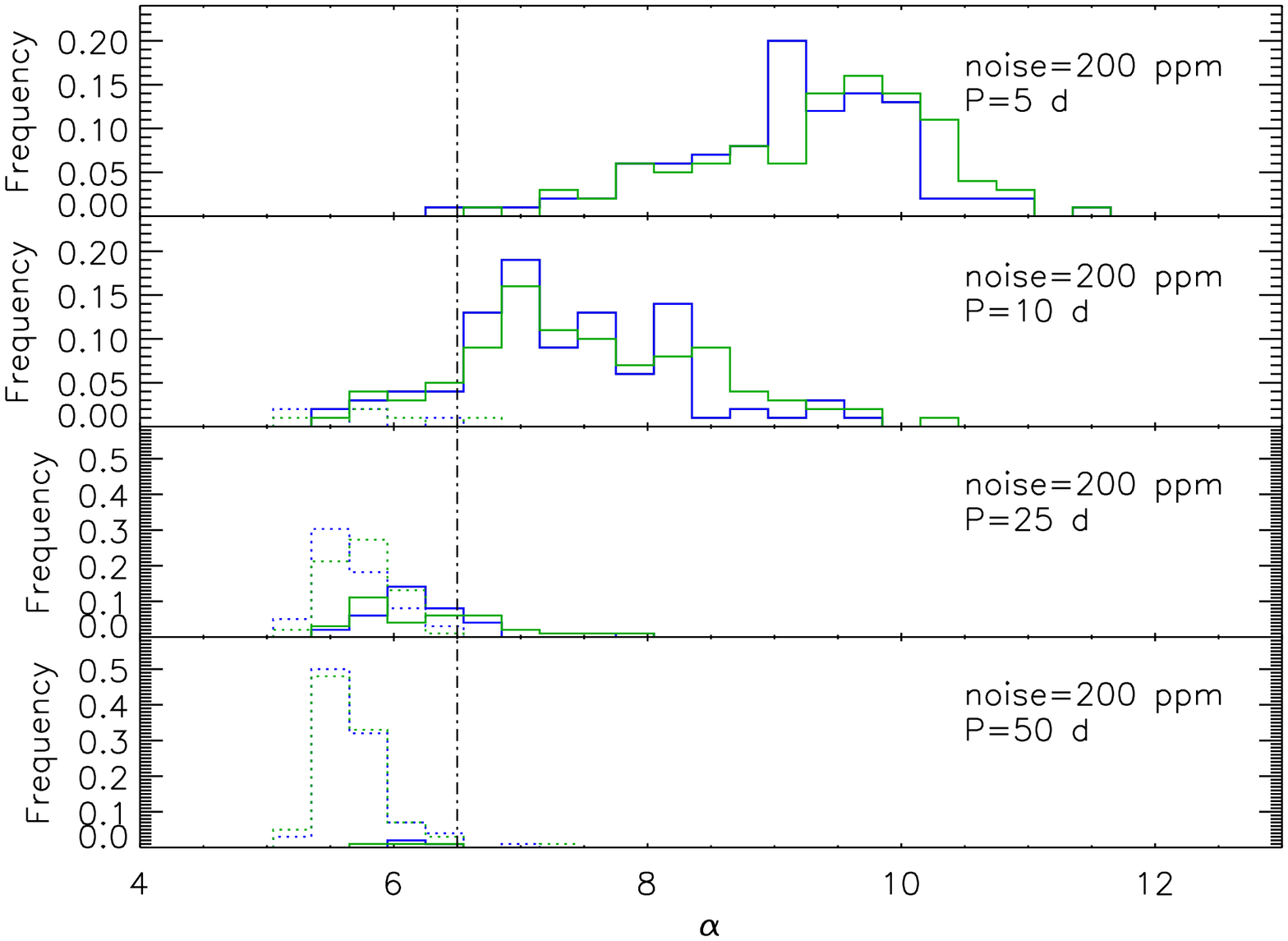} 
% \vspace*{-1.0 cm}
\end{center}
\end{minipage}
}
\begin{center}
\caption{\emph{Left}: Distributions of the values of $\alpha$, the signal-to-noise ratio of a transit detection, obtained by analysing light curves with transits of a planet of 1.5 R$_{\oplus}$. Each set is characterized by a different orbital period of the planet (as labelled). The standard deviation of the white Gaussian noise is in all cases 200 ppm. The red and blue vertical dot-dashed lines indicate the 1 percent false-alarm threshold for the 200-harmonic fitting and the 3-spot model, respectively. Red solid histograms show the statistics of light curves where the period $P$ was correctly identified after applying the 200-harmonic method, blue solid histograms show those after applying the 3-spot method. Dashed histograms refer to the statistics of light curves where the period $P$ was incorrectly identified, with the same color coding. 
\emph{Right}: As on the left, distributions of the values of $\alpha$ obtained analysing the simulated light curves after applying the 3-spot model (blue histograms) and the INL filter (green histograms).}
   \label{fig1}
\end{center}
\end{figure}

\section{Results and conclusions}
\noindent
For $\sigma= 100$ ppm, the filtering methods that achieve the best performance are the INL filter and the 200-harmonic fitting with detections up to 98 percent for $R=1.25$ R$_{\oplus}$ and $P=10$ days. In most of the cases they give comparable results, although in some instances the INL filter has a slightly better performance, owing to the Gibbs phenomenon (\cite[Morse \& Feshbach 1954]{MorseFeshbach54}) affecting the 200-harmonic fitting.

\begin{figure}[!b]
 \vspace*{-1.0 cm}
\begin{center}
 \includegraphics[width=6cm,angle=90]{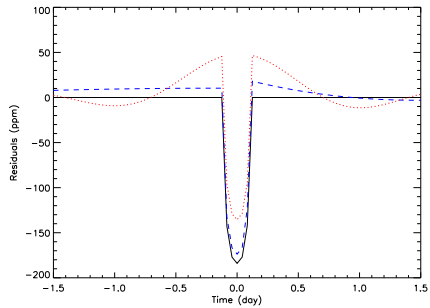} 
\vspace*{-0.8 cm}
 \caption{The shape of a transit of an Earth-like planet as it appears in the ideal case (solid black line), in the residuals obtained with the 3-spot model (dashed blue line) and in those of the 200-harmonic fitting (dotted red line). Note the reduction of the transit depth and the overshooting at the edges of the transit dip due to the Gibbs phenomenon in the case of the 200-harmonic fitting.}
\label{fig2}
\end{center}
\end{figure}

When $\sigma \geq 200$ ppm, the method with the best performance is the INL filter when we use an appropriate window of 2 days for the median boxcar filter. It shows a performance comparable with that of the 3-spot model in most of the cases, even better in some instances (see Fig.~\ref{fig1}). On the other hand, the 200-harmonic fitting has the worst performance because of the Gibbs phenomenon (see Fig.~\ref{fig1} and ~\ref{fig2}).

\begin{figure}[!t]
% \vspace*{-2.0 cm}
\begin{center}
 \includegraphics[width=8cm]{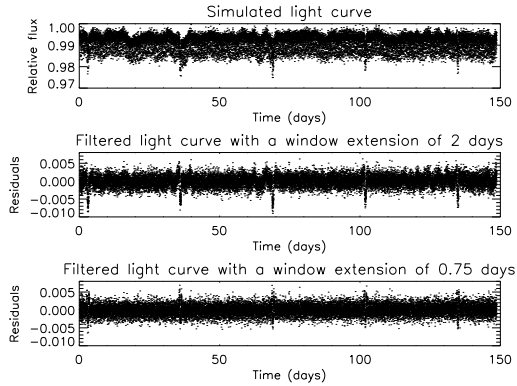} 
 \vspace*{-0.4 cm}
 \caption{\emph{Upper panel}: one of the light curves with transits simulated for the First CoRoT Blind test (\cite[Moutou et al. 2005]{Moutouetal2005}; ID=460). \emph{Middle panel}: the light curve filtered by means of the INL filter with a window of 2 days. \emph{Bottom panel}: the filtered light curve with a 0.75 day window. Note the disappearance of the transits when the window extension is reduced.}
  \label{fig3}
\end{center}
\end{figure}

The performance of the INL filter depends critically on the adopted extension of the filter window. An optimal window of 2 days has been chosen for our analysis. Shorter windows negatively affect the transit detection since they give rise to a reduction of the transit depth in the filtered light curve (see Fig.~\ref{fig3}), in which case the 3-spot model and the 200-harmonic fitting would prove to be the best methods for the cases with $\sigma \geq 200$ ppm and $\sigma= 100$ ppm, respectively.

The optimal width of the median filter window depends on the magnetic activity level of the star and its rotation period. Specifically, the higher the activity level and the shorter the rotation period, the shorter the optimal window, because the time scales of the flux variations decrease with increasing activity. In other words, in the case of highly active stars the window extension has to be shortened with respect to the solar case, otherwise some oscillations or transit-like features will appear in the residuals owing to a bad filtering of the variability. To fix automatically the window extension, we propose a method similar to that of \cite[Regulo et al. (2007)]{Reguloetal2007}, computing the power spectrum of the time series and choosing an extension corresponding to the frequency where the power spectral density goes below a fixed threshold, usually set at $10^{-6}$ of the maximum power level.

We conclude that the INL filter, when applied with a suitable choice of its window, has a better performance than more complicated and computationally intensive methods of fitting solar-like variability, like the 200-harmonic fitting or the 3-spot model.

\begin{acknowledgements}
\noindent
The authors gratefully acknowledge support from the Italian Space Agency (ASI) under contract ASI/INAF I/015/07/0, work package 3170. This research has made use of results produced by the PI2S2 Project managed by the Consorzio COMETA, a project co-funded by the Italian Ministry of University and Research (MUR) within the {\it Piano Operativo Nazionale "Ricerca Scientifica, Sviluppo Tecnologico, Alta Formazione" (PON 2000-2006)}.
\end{acknowledgements}

\end{document}